\documentclass[reprint,aps,pra,showpacs,floatfix]{revtex4-1}
\usepackage{graphics,epsfig,graphicx}
\usepackage{amsbsy}
\usepackage{amsmath}
\usepackage{bm}
\usepackage{amsfonts}
\usepackage{cancel}
\usepackage{multirow}
\begin{document}

\title{Universal range corrections to the Efimov trimer for a class of
paths to the unitary limit}

\author{A. Kievsky} 
\affiliation{Istituto Nazionale di Fisica Nucleare, Largo Pontecorvo 3, 56100 Pisa, Italy}
\author{M. Gattobigio}
\affiliation{Universit\'e de Nice-Sophia Antipolis, Institut Non-Lin\'eaire de
Nice,  CNRS, 1361 route des Lucioles, 06560 Valbonne, France }

\begin{abstract}
Using potential models we analyze range corrections to the universal law
dictated by the Efimov theory of three bosons. In the case of finite-range
interactions we have observed that, at first order, it is necessary to supplement
the theory with one finite-range parameter, $\Gamma_n^3$, for each specific
$n$-level [Kievsky and Gattobigio, Phys. Rev. A {\bf 87}, 052719 (2013)].  
The value of $\Gamma_n^3$  depends on the way the potentials is 
changed to tune the scattering length toward the unitary limit. In this
work we analyze 
a particular path in
which the length $r_B=a-a_B$, measuring the difference between the two-body
scattering length $a$ and the energy scattering length $a_B$, results almost
constant. Analyzing systems with very different scales, as atomic or nuclear
systems, we observe that the finite-range parameter remains almost
constant along the path with a numerical value of $\Gamma_0^3\approx 0.87$ for
the ground state level. This observation suggests the possibility of
constructing a single universal function that incorporate finite-range effects
for this class of paths. The result is used to estimate the three-body
parameter $\kappa_*$ in the case of real atomic systems brought to the unitary
limit thought a broad 
Feshbach resonances.
Furthermore, we show that the finite-range parameter can be put in relation with the
two-body contact $C_2$ at the unitary limit.  
\end{abstract}
\maketitle

\section{Introduction}

The study of universal behavior in few-boson systems is an intense subject of research
nowadays. Universal properties appear for example in loosely bound systems in
which the particles stay most of the time outside the range of the interaction. 
In this situation the details of the interaction between components are not important
and the dynamics of the system can be described in terms of few control
parameters. In the two-body system the scattering length $a$ can be used as the control parameter. 
In fact, when the
two-body binding energy $E_2\rightarrow 0$ its value results $E_2\approx \hbar^2/ma^2$
and all of the two-body observables can be written in terms $a$ as
well~\cite{braaten:2006_physicsreports}. The paradigm for this universal
behavior is represented by the
zero-range theory: the particles are outside the interaction range all of the time.
In this case the above relation becomes exact, $E_2 = \hbar^2/ma^2$, and the
usual effective range expansion for the $s$-wave phase-shift reduces to
$k\cot\delta=-1/a$ (with the total energy $E=\hbar^2 k^2/m$).

Application of this theory to the three-boson system produces the Thomas
collapse~\cite{thomas:1935_phys.rev.}: the ground state energy is unbounded from below. 
Besides this singular behavior, the three-boson system shows a very
peculiar behavior as the two-body scattering length approaches the
unitary limit, $1/a\rightarrow 0$. In this limit, as has been shown by V. Efimov in a series of 
papers~\cite{efimov:1970_phys.lett.b,efimov:1971_sov.j.nucl.phys.}, a system of
three identical bosons interacting through a two-body short-range interaction
shows a geometrical series of bound states whose energies accumulate to zero. 
The ratio between the energies of two consecutive states is constant and
does not depend on the nature of the interaction. This particular behavior is
known as the Efimov effect and its observation has triggered an
enormous amount of experimental as well as theoretical work in 
different fields as molecular, atomic, nuclear and particle physics 
(for instance see Refs.~\cite{ferlaino:2011_few-bodysyst.,frederico:2011_few-bodysyst.} 
and references therein).

The spectrum of the three-boson system close to the unitary limit is described
by the Efimov equation (or Efimov radial law) which can be expressed in a
parametric form as follow
\begin{subequations}
  \begin{eqnarray}
    \label{eq:energyzrA}
      E_3^n/(\hbar^2/m a^2) = \tan^2\xi \\
      \kappa_*a = \text{e}^{(n-n^*)\pi/s_0} 
      \frac{\text{e}^{-\Delta(\xi)/2s_0}}{\cos\xi}\,,
    \label{eq:energyzrB}
  \end{eqnarray}
    \label{eq:energyzr}
\end{subequations}
with $\Delta(\xi)$ a universal function whose parametrization can be
found in Ref.~\cite{braaten:2006_physicsreports} 
and $s_0\approx 1.00624$ is a universal number. 
The scale at which a particular set of eigenvalues are selected among the
infinite set of values is fixed by
$\kappa_*$, called the three-body parameter, which defines the energy 
$\hbar^2 \kappa_*^2/m$ for $n=n^*$ at the unitary limit. Knowing the
value of $\kappa_*$ the spectrum in terms of $a$ is completely determined.

The ability of tuning $a$ in atomic-trapped systems has allowed different
experimental groups to measure the value of the scattering length $a_-$ at which
the three-body bound state disappears into the continuum ($\xi\rightarrow
-\pi$). From Eq.~({\ref{eq:energyzr}) we can see that measuring $a_-= -
\text{e}^{-\Delta(\pi)/2s_0}/\kappa_*\approx -1.50763/\kappa_*$, is an indirect
way for measuring the three-body parameter $\kappa_*$. As an interesting result,
it has been experimental
found~\cite{ferlaino:2011_few-bodysyst.,machtey:2012_phys.rev.lett.,
roy:2013_phys.rev.lett.,dyke:2013_phys.rev.a}, and theoretically
justified~\cite{wang:2012_phys.rev.lett.,naidon:2014_physrevA},
that in the class of alkali atoms $a_-/\ell\approx -9.5$, with $\ell$ the van
der Waals length. Recently the same behavior has been seen in a gas of $\,^4$He
atoms~\cite{knoop:2012_phys.rev.a}.  This fact has extended the discussion of
universal behavior to analyzed the dependence of the three-body parameter on the
two-body dynamics~\cite{naidon:2012_phys.rev.a}.

In the above discussion we have used that
Eq.~({\ref{eq:energyzr}) and the parametrization of $\Delta(\xi)$
have been derived in the zero-range limit (scaling limit). 
On the other hand experiments and calculations made for real systems deal with 
finite-range interactions and, for this reason, finite-range 
corrections have to be considered~\cite{ji:2010_epl}.
In Refs.~\cite{kievsky:2013_phys.rev.a,gattobigio:2014_phys.rev.a,kievsky:2014_phys.rev.a}, 
the authors have 
solved the Schr\"odinger equation using potential models in order to 
observe in which manner finite-range corrections manifest in numerical
calculations. The following modifications to the Efimov radial law
have been proposed
\begin{subequations}
  \begin{eqnarray}
  \label{eq:energyfrA}
    E_3^n/E_2 = \tan^2\xi \\
    \kappa^3_na_B + \Gamma_n^3 =
    \frac{\text{e}^{-\Delta(\xi)/2s_0}}{\cos\xi} \,.
  \label{eq:energyfrB}
  \end{eqnarray}
  \label{eq:energyfr}
\end{subequations}

In the above equation there are two different corrections: one comes from the two-body 
sector and is
taken into account by substituting $a_B$ for $a$, defined by
$E_2=\hbar^2/m a_B^2$, with $E_2$ the two-body binding energy if $a>0$, or the two-body
virtual-state energy in the opposite case, $a<0$~\cite{ma:1953_rev.mod.phys.}.
The second correction enters as a shift $\Gamma_n^3$ in the control parameter $\kappa_*a_B$. 
As we will show below the shift is almost constant close to the unitary limit.
Moreover, the values of the three-body parameter and the shift depend on the energy level.

In both Eqs.~(\ref{eq:energyzr}) and (\ref{eq:energyfr}) the three-body energy
is expressed as a function of the scattering length. In principle, there can be
different ways and different mechanisms to change the scattering length and tune
it to $a\rightarrow\pm\infty$, and we refer to a given protocol as a path to the
unitary limit. In this work we explore one of these paths, or one class of
paths, in which the length $r_B = a - a_B$ is kept constant.
Along this path we obtain that the shift $\Gamma_n^3$, which takes into account
range corrections in the three-body sector, is independent, at a first order, on 
the kind of potential we use. This fact points to a kind of universality of the 
first-order range corrections. 

One can speculate that this class of paths to the unitary limit is pertinent for
the description of broad Feshbach resonances; using the universal corrections we
can thus estimate the value of the 3-body parameter $\kappa_*$ and compare it to
what found in literature. 
Finally, we show that the shift $\Gamma_0^3$ can be related to the two-body
contact $C_2$ for a trimer at the unitary limit. 

The paper is organized as follow: in the Sec.~\ref{sec:path} we study the
path to the unitary limit which describes a broad Feshbach resonance, 
and we analyze the behaviour of the two-body properties, in particular of 
$r_B$.
In Sec.~\ref{sec:threeBody} the three-body system is studied along that path
using different potential models. We construct the Gaussian universal function,
and we argue on its universality inside the potential models.  In particular we
show how to use it to predict the three-body parameter $\kappa_*$ in real
systems. In the
same section we show the relation between the two-body contact and the shift.
In Sec.~\ref{sec:conclusions} we make our conclusions.

\section{Path to the unitary limit}\label{sec:path}
In the zero-range Efimov theory $a_B=a$ and this is the only 
way to move toward the unitary limit. On the
contrary, in a potential-based theory 
the finite-range character of the interaction allows for different paths
connecting the physical point to the unitary limit. 
For example, in atomic-trapped systems, experimentalists use Feshbach resonance to
modified the interatomic potential. By changing the intensity of the magnetic field the two-body
scattering length $a$ is modified and for particular values of the magnetic field 
it diverges. These resonances, which are further
classified as broad or narrow, can be interpreted as a particular path along
which the unitary limit is reached. Different theoretical descriptions of
Feshbach resonances are available and the most common describes this
process as a coupled channel system~\cite{chin:2010_rev.mod.phys.}; still, broad
resonances can be simply modeled by modifying the potential strength. In
the following we analyze this option. We define
\begin{equation}
V_\lambda(r)=\lambda V(r)\,,
\label{eq:potl}
\end{equation}
a potential with variable strength where $V(r)$ is the potential that reproduces
the binding energy $E_2$ and the two-body scattering length $a$ of a particular system.
The original potential corresponds to $\lambda=1$, and the unitary limit is
reached decreasing the value of $\lambda$ down to a critical value $\lambda_c$.
Examples could be a dimer of two helium atoms, $E_2\approx 1.3\,$mK and
$a\approx 190\,$a$_0$,
or the deuteron, $E_2\approx 2.22\,$MeV and $a\approx 5.4\,$fm (here $a$ is the triplet scattering
length). To analyze these two cases we consider the LM2M2 potential
of Aziz~\cite{aziz:1991_j.chem.phys.} for the helium dimer and, in the case of the deuteron, 
a combination of two Yukawians as the MTIII nucleon-nucleon 
potential~\cite{malfliet:1969_nuclearphysicsa}. It should be noticed that these
potentials reproduced a large set of two-body data and, for this
reason, are considered realistic potentials. On the other hand the Efimov radial
law depends only on $a$ or $a_B$, this suggests the use of this minimal
information to construct a two-parameter potential as a Gaussian
\begin{equation}
V(r)=V_0 e^{-r^2/r_0^2}
\label{eq:potg}
\end{equation}
with the strength and range fixed to describe the two experimental data. In the
specific cases analyzed here the LM2M2 and MTIII interactions
can be mimicked by a Gaussian of range $r_0=r_0^{\rm He}=10$ a$_0$ and 
range $r_0=r_0^{NN}=1.65$ fm, respectively. 

Using the four different potential models, we have studied the behavior of the
energy-scattering length $a_B$,  of the scattering length $a$, and of the
effective range $r_\text{eff}$ close to $\lambda_c$. In this region these two
lengths are related by the effective range expansion as
\begin{equation}
\frac{1}{a_B}\approx\frac{1}{a}+\frac{r_\text{eff}}{2a_B^2} \, .
\label{eq:effect_ab}
\end{equation}
If we define $\epsilon = \lambda-\lambda_c$, we 
have $a,a_B = {\cal O}(1/\epsilon)$, while  
$r_\text{eff} = r_u + {\cal O}(\epsilon)$, or $r_\text{eff} = r_u + {\cal O}(1/a)$, if we want
to emphasize the dependency on the scattering length. For convenience we have
introduced the length $r_u$ as the value of $r_\text{eff}$ at the
unitary limit. Formally $r_u$ can be obtained from the effective range expansion
that, at that limit, takes the form
\begin{equation}
\lim_{a\rightarrow\infty} \frac{\cot\delta_0}{k} = \frac{1}{2}r_u-Pr_u^3k^2 + \cdots  \, 
\label{eq:effect_ru}
\end{equation}
where $\delta_0$ is the $s$-wave phase-shift, $k$ is the energy momentum 
and $P$ is the shape parameter. Therefore
\begin{equation}
r_u=\lim_{a\rightarrow\infty} r_\text{eff} 
 =2\lim_{k\rightarrow 0} \left[\lim_{a\rightarrow\infty}
\frac{\cot\delta_0}{k}\right] \,\, .
\label{eq:limit_ru}
\end{equation}

What we have observed is that while $r_\text{eff}$ changes
considerably from its value $r_u$ as the system moves away from 
the unitary limit, the length $r_B$ defined as
\begin{equation}
  r_B = a-a_B\,,
  \label{eq:rb}
\end{equation}
results almost constant, even if $r_B = r_u/2 + {\cal O}(1/a)$.
To better analyze this fact we write
Eq.(\ref{eq:effect_ab}) in the form
\begin{equation}
  \frac{r_\text{eff}}{2r_B}=1-\frac{r_B}{a} + {\cal O}(1/a^2)\,.
  \label{eq:rbvsre}
\end{equation}
If $r_B$ were strictly constant, we could replace in the above equation
$2r_B$ by $r_u$, its value at the unitary limit, resulting in an universal
relation for the ratio $r_\text{eff}/r_u$ in terms of the variable
$r_u/a$
\begin{equation}
  \frac{r_\text{eff}}{r_u}=1-0.5\frac{r_u}{a} \,.
  \label{eq:rbvsren}
\end{equation}
In Fig.~\ref{fig:rb} we show $r_\text{eff}/{r_u}$ as a function
of the inverse scattering length for different potential models. 
We can observe that these very different potentials, describing
physical system with different scales as the atomic LM2M2 or 
the nuclear MTIII interactions collapse on a very narrow
band; this is more evident for positive values of $a$ where the physical points
are located. The numerical results for the different values of $\lambda$
are given by the solid points whereas the solid lines 
are fits to the numerical results using a linear plus a quadratic $r_u/a$ term. 
The results for the two Gaussian potentials, given by the solid (green) circles
lie on a single line.

The behavior of $r_B$ is also shown in Fig.~\ref{fig:rb}, 
in the case of the Gaussian interactions,
by the (red) circles. The (red) dashed line is a fit to the numerical
results using a linear plus a quadratic $r_u/a$ term. It should be noticed
that in the expansion of $2r_B/r_u\approx 1+{\cal A}_0r_u/a+\ldots$, the
coefficient of the linear term, ${\cal A}_0 \approx -0.01$  
is more than one order of magnitude smaller than the corresponding one in 
the expansion of $r_\text{eff}/{r_u}$ (which results to be always very
close to 0.5) resulting in the almost constant behavior of $r_B$.
In the figure the position of the dimer formed by two helium atoms and
the deuteron are explicitly shown.

\begin{figure}[h]
\vspace{0.8cm}
\begin{center}
\includegraphics[width=\linewidth]{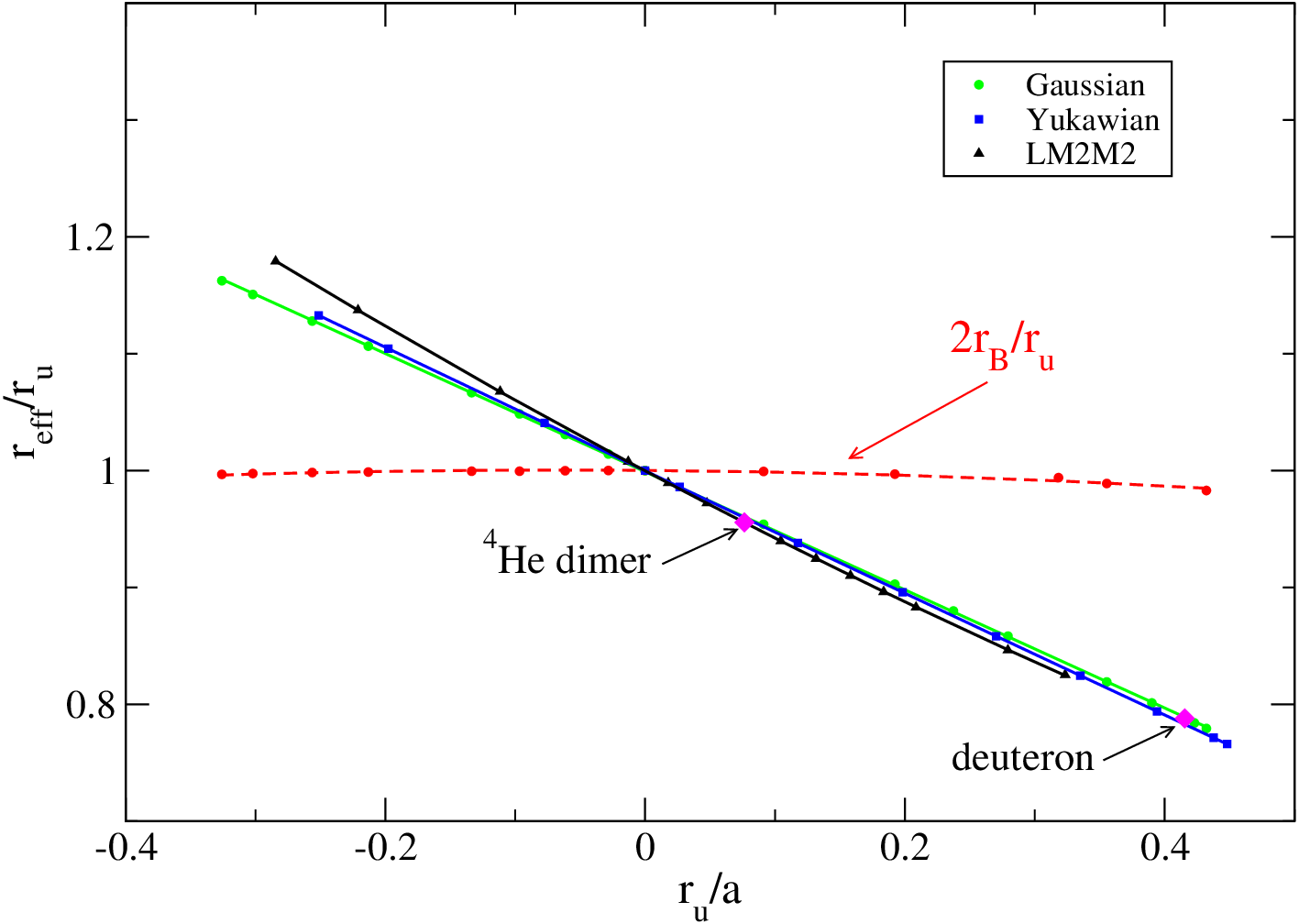}
\end{center}
\caption{(color online). The effective range $r_\text{eff}$ as a function of the
inverse of the scattering length (in units of $r_u$)
for different potential models. The length $2r_B$ in units of $r_u$ is also
shown.} 
\label{fig:rb}
\end{figure}

\section{The three-body sector}\label{sec:threeBody}

The modifications to the Efimov radial law proposed in Eq.(\ref{eq:energyfr})
have been derived using potential models following the path to the unitary limit discussed
above. The potential strength has been varied in order to cover a wide range
of values of $a$ from positive to negative values up to $a_-$, the value at which the
three-body system disappears into the three-body continuum. 

Let concentrate our analysis to the ground state of the three-body system, that
means $n=n^*=0$, such that $\kappa_*=\kappa^3_0$;  in this case
Eq.~(\ref{eq:energyzrB}) and Eq.~(\ref{eq:energyfrB}) are related by 
\begin{equation}
a(\xi)-a_B(\xi)=\Gamma_0^3/\kappa_* \,,
\end{equation}
where we want to stress that  $a$ and $a_B$ are evaluated at the same angle $\xi$ 
and so
they are not related to the same $E_2$ value but to the respective solutions of
Eqs.~(\ref{eq:energyzrA}) and (\ref{eq:energyfrA}). The above relation extends the
concept of almost constant behavior of the difference $a-a_B$ from the two-body
sector to the three-body sector. 
Defining $r_*=\Gamma_0^3/\kappa_*$, this length
represents the distance between the zero-range and the finite-range theory
in the $(1/a,-\sqrt{|E^0_3|})$ plane and, if measured at fixed $\xi$ angles,
it results (almost) constant.
In order to further analyze this fact we discuss first the case of a Gaussian
potential. 

\subsection{Gaussian universal function}

The three-boson ground-state energy, $E_3^{0,G}$, has been calculated 
with the two Gaussian potentials for different values of $\lambda$.
At $\lambda=\lambda_c$ the results verify
\begin{equation}
\kappa_*^{\rm He} r_0^{\rm He}= \kappa_*^{NN} r_0^{NN}\approx 0.488
\label{eq:0.488}
\end{equation}
where $\kappa_*^{\rm He}$ and $\kappa_*^{NN}$ are obtained from the
energy at the unitary limit in both cases; thus, for a Gaussian
of range $r^G_0$ the ground-state binding energy at the unitary limit is
$E_*^G\approx \hbar^2(r^G_0/0.488)^2/m$. 
The ground state energies close to the unitary
limit can be analyzed using Eq.(\ref{eq:energyfr}). The results can be described
with high accuracy using a shift $\Gamma_0^3$ having a constant plus a $1/(\kappa_* a_B)$ 
term
\begin{equation}
\Gamma_0^3 = \Gamma_{0,0}^3 + \frac{\Gamma_{0,1}^3}{\kappa_*a_B}\, .
\label{eq:expg}
\end{equation}
The parameters $\Gamma_{0,0}^3\approx 0.87$ and $\Gamma_{0,1}^3\approx -0.14$
are independent of the range of the Gaussian
potential. The small value of $\Gamma_{0,1}^3$  guarantees an
almost constant behavior of the shift along the path of variable strength.
This behavior 
can be explicitly studied 
by casting 
Eq.~(\ref{eq:energyfrB}) in the following form
\begin{equation}
    1+\frac{\Gamma^3_0}{\kappa_*a_B} = 
      \frac{\text{e}^{-\Delta(\xi)/2s_0}}{\kappa_*a_B\cos\xi} \,.
  \label{eq:eqgammac}
\end{equation}
The right hand side can be plotted against $1/\kappa_*a_B$ looking for
a linear behavior. This is shown in Fig.~\ref{fig:gammac} where the
solid points represent the calculations with the Gaussian potentials and
the solid line is a fit using the expansion of Eq.(\ref{eq:expg}).
From the figure the almost constant behavior of $\Gamma_0^3$, close to
the unitary limit, is confirmed. Moreover this analysis shows the completely
equivalence between the different Gaussian potentials using $\kappa_*$
as a scale factor.

\begin{figure}[h]
\vspace{0.8cm}
    \begin{center}
       \includegraphics[width=\linewidth]{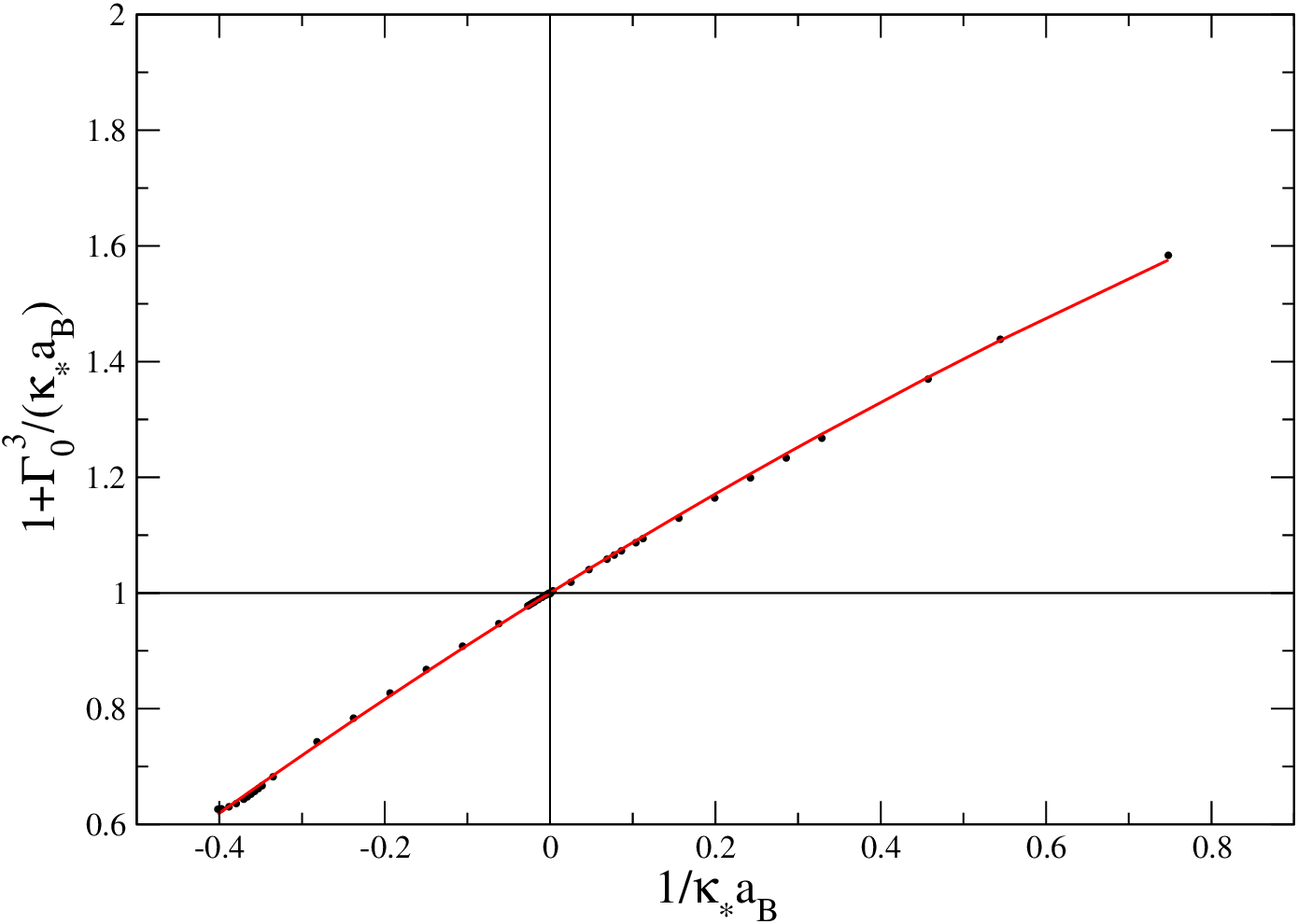}
    \end{center}
    \caption{(color online).  Study of Eq.(\ref{eq:eqgammac}) as a function of
$1/\kappa_*a_B$. The solid circles are the numerical results for the r.h.s. of
the equation. The (red) solid line is a fit to the calculations using a linear
plus quadratic term.} 
  \label{fig:gammac}
  \end{figure}

The equivalence between Gaussian potentials with 
different ranges suggests the possibility of defining
a Gaussian universal function $\widetilde\Delta(\xi)$. The
modified radial law of Eq.(\ref{eq:energyfr}) can be written as
\begin{subequations}
  \begin{eqnarray}
    E_3^0/E_2 = \tan^2\xi \\
    \kappa_*a_B = \frac{\text{e}^{-\widetilde\Delta(\xi)/2s_0}}{\cos\xi} \,,
  \label{eq:energyfrGB}
  \end{eqnarray}
  \label{eq:energyfrG}
\end{subequations}
and $\widetilde\Delta(\xi)$
can be extracted from the numerical solutions of the ground state energy,
$E^0_3$, of the three-boson system calculated with a Gaussian interaction as
\begin{equation}
\widetilde\Delta(\xi) = s_0\; {\rm ln}
\left(\frac{E^0_3+E_2}{\hbar^2\kappa_*^2/m}\right) \, .
\end{equation}

The Gaussian universal function calculated in this way incorporates finite-range
effects along the specific path used
to reach the unitary limit. We would like to stress that exists only one
Gaussian universal function independently of the range and strength of the
Gaussian used to calculate it.
In Fig.~\ref{fig:uni1} the Gaussian function $\widetilde\Delta(\xi)$ 
calculated using the two Gaussian potentials is given by the solid (green)
circles. The two set of data completely overlap and cannot be distinguished
from which Gaussian potential they have been calculated. The Gaussian universal function
is compared to calculations using the LM2M2 interaction, (blue)
squares, and the Yukawian MTIII potential, (black) triangles. 
For the sake of comparison the zero-range universal function $\Delta(\xi)$
is shown in the figure by the solid (red) line.

Interestingly, these interaction models, very different in scale and functional
form, give rise to equivalent functions that collapse in a narrow band
indicating that they have similar $\Gamma^3_{0,0}$ values. In fact for the LM2M2
potential the value $\Gamma^3_{0,0}\approx 0.82$ is obtained whereas for the MTIII potential 
$\Gamma^3_{0,0}\approx 0.85$. This means that, up to first order, the range
corrections close to the unitary limit are almost universal. Moreover,
we can conclude that close to the unitary limit a two-parameter
potential as a Gaussian contains the essential ingredients required by the
dynamics. The different 
scales (here we have explored atomic and nuclear scales) are absorbed in the 
control parameter $\kappa_*a_B$. The extension of the analysis to the first
excited state, $n=1$ level, produce a shift with a constant term of
$\Gamma^3_{1,0}\approx0.08$ for the potentials under consideration.

\begin{figure}[h]
\vspace{0.8cm}
    \begin{center}
      \includegraphics[width=\linewidth]{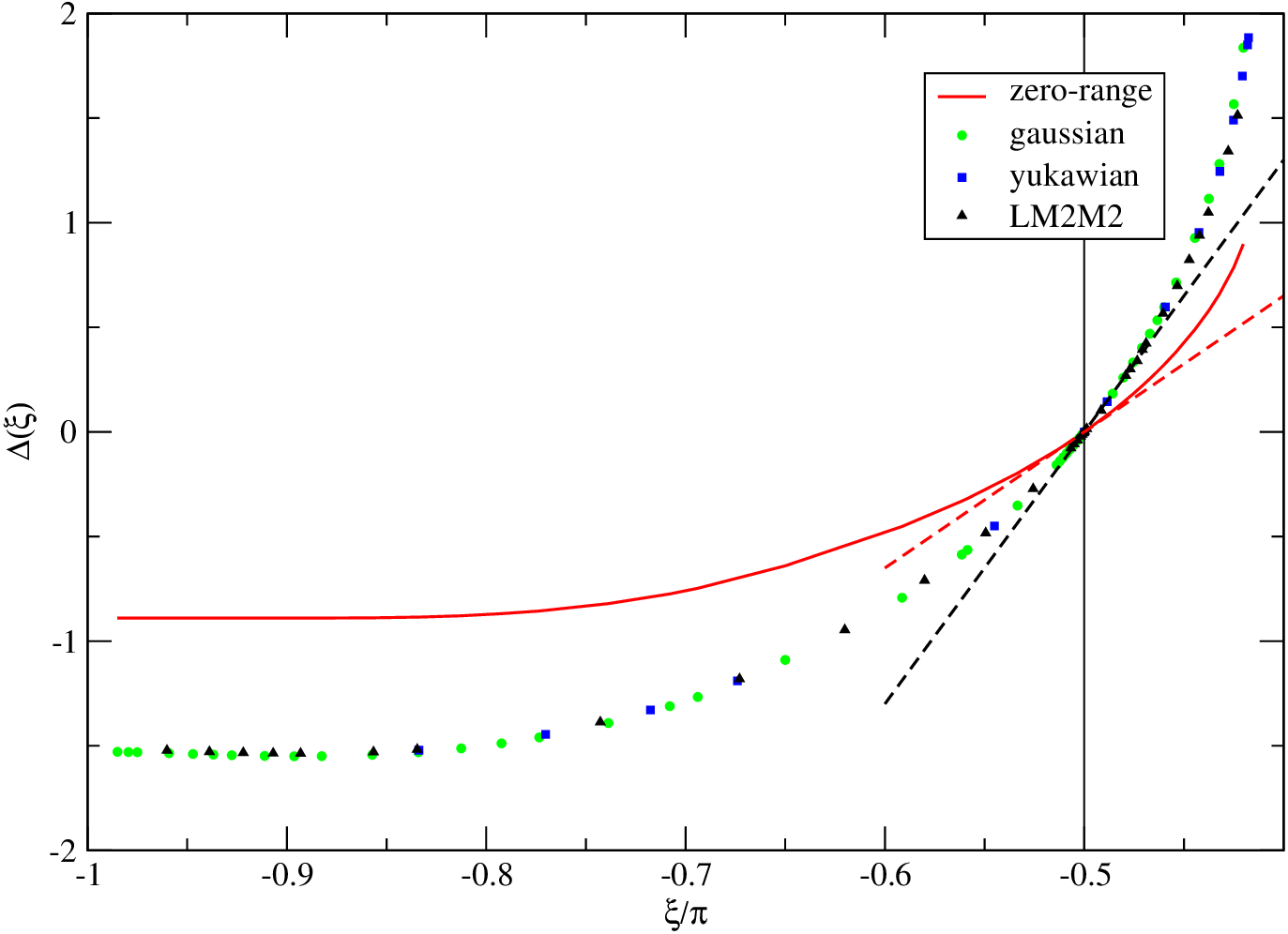}
    \end{center}
    \caption{(color online). The zero-range universal function, (red) solid
line, as
a function of the angle $\xi$. The (green) circles, (blue) squares and
(black) triangles are the equivalent functions calculated for the corresponding
potentials. The dashed lines are the tangents at the unitary limit.} 
  \label{fig:uni1}
  \end{figure}

  \subsection{Estimation of $\kappa_*$}
Although the equivalent functions are very similar for the Gaussian potential
and for the realistic ones, differences are evident when these potential are
used to calculate energy levels in the three-boson system. The Gaussian adapted
to reproduce the values of the LM2M2 interaction in the two-body system has to
be supplemented with a three-body force in order to reproduce the corresponding
energies in the three-boson system. For example, in 
Ref.~\cite{kievsky:2011_few-bodysyst.} the low energy dynamics
of three helium atoms has been described using a soft potential model
consisting in a two-body Gaussian plus a Gaussian hypercentral force adapted 
from the LM2M2 interaction. Moving from the physical point towards the
unitary limit this strategy implies that a set of calculations have to be performed in order
to adapted the soft potential model to the realistic one. 

Instead of this
type of approach here we want to use the Gaussian universal function 
to estimate the three-body parameter, $\kappa_*$, knowing $a$, $E_2$ and $E^0_3$ 
for some specific system. This can be done observing that
Eq.(\ref{eq:energyfrG}) is a one-parameter equation as Eq.(\ref{eq:energyzr}).
The finite-range effects has been absorbed in $\widetilde\Delta(\xi)$. 
An operative way to obtain this estimate is to determine
the range and strength of the Gaussian from $a$ and $E_2$. 
Calling $r_0^G$ the range determined in this way,
the three-body parameter for that Gaussian can be determined from the relation 
of Eq.(\ref{eq:0.488}), $\kappa_*^G=0.488/r_0^G$. 
Finally the three-body parameter of the system
can be obtained from the following relation derived using Eq.(\ref{eq:energyfrG}) at
a fixed value of the angle $\xi$ 
\begin{equation}
   \kappa_* \approx\kappa_*^G \sqrt{\frac{E^0_3}{E^{0,G}_3}}
\label{eq:kestimate}
\end{equation}
where $E^{0,G}_3$ is three-boson ground state energy calculated 
using the Gaussian potential at a strength value such that
$E^0_3/E_2=E^{0,G}_3/E^G_2$ and with $E^G_2$ the two-body energy
calculated at that strength. 

For example, using the LM2M2 interaction the 
helium dimer and trimer energies are $E_2=1.303\;$mK and $E^0_3=126.4\;$mK 
respectively, and $E^0_3/E_2=\tan\xi^2=97.0$. Using the Gaussian representation with
$r_0^G=r_0^{\rm He}=10$a$_0$ and a strength $V_0=1.24294$K the following values are
obtained: $E^G_2=1.6209\;$mK and $E^{0,G}_3=157.3 \;$mK, verifying $E^0_3/E_2=\tan\xi^2=97.0$.
Using the above equation, we obtain an estimate of the three-body parameter 
$\kappa_*\approx 0.0437$a$_0^{-1}$ to
be compared to the value of $0.0440$a$_0^{-1}$ obtained calculating the
three-boson ground state energy with the LM2M2 potential using Eq.(\ref{eq:potl})
with $\lambda_c=0.9743$. In the case of the MTIII potential a similar reasoning 
produces and estimate of the three-body parameter of
$\kappa_*\approx 0.269$fm$^{-1}$ compare to the exact calculation giving 
a value of $0.277\;$fm$^{-1}$. 
This results are well below a $3\%$ accuracy, and a similar accuracy has been
obtained for example using a next to the leading order effective field theory
description~\cite{braaten:2003_phys.rev.a}.

\subsection{Finite-range parameter in terms of the contact}
The shift modifies the Efimov radial law equation to take into account
finite-range corrections~\cite{gattobigio:2014_phys.rev.a}; thus, we have
shown that the description of an Efimov-state energy requires two 
three-body parameters, $\kappa_*$ and the shift. We can relate the 
two parameters to other properties of a bosonic gas at the unitary limit: the two- and
three-body
contacts~\cite{tan:2008_annalsofphysics,tan:2008_annalsofphysics-1,tan:2008_annalsofphysics-2}.

We briefly recall that the two- and three-body contacts are defined by the following
relations
\begin{subequations}
  \begin{eqnarray}
  \label{eq:contact2}
  \left(a\frac{\partial E}{\partial a}\right)_{\kappa_*} 
  = \frac{\hbar^2}{8\pi m a} C_2\,,   \\
  \left(\kappa_*\frac{\partial E}{\partial \kappa_*}\right)_a
  = -\frac{2\hbar^2}{m} C_3\,.   
  \label{eq:contact3}
  \end{eqnarray}
  \label{eq:contacts}
\end{subequations}
They are important quantities entering in several
properties of a many body system. For instance, the tail of the momentum
distribution $n(k)$ of a bosonic gas contains $C_2$ and $C_3$ in its asymptotic behaviour 
\begin{equation}
  n(k) \rightarrow \frac{1}{k^4} C_2 + \frac{F(k)}{k^5} C_3\,.
  \label{eq:nofk}
\end{equation}
In Eq.~(\ref{eq:nofk}) $F(k)$ is an universal log-periodic
function~\cite{castin:2011_phys.rev.a,braaten:2011_phys.rev.lett.,smith:2014_phys.rev.lett.},
it is not relevant for what follows.

In Refs.~{\cite{werner:2010_,castin:2011_phys.rev.a}} $C_2$ has been calculated for 
the Efimov trimer in the unitary limit, $a\rightarrow \pm\infty$,
and the result can be expressed in the following  way
\begin{equation}
  C_2 = 8\pi\frac{\Delta'(-\pi/2)}{s_0}\,\kappa_* \approx 53.01\,\kappa_*\,,
  \label{eq:c2Universal}
\end{equation}
where $\Delta'(-\pi/2)\approx 2.125850$~\cite{werner:2010_} is the derivative of
the universal function calculated at the unitary point.
In the same way, we can consider the two-body contact for the ground-state 
trimer calculated with the Gaussian potential; using Eq.~(\ref{eq:energyfrG}) 
we obtain 
\begin{equation}
  \widetilde C_2 = 8\pi\frac{\widetilde\Delta'(-\pi/2)}{s_0}\,\kappa_*
  \approx 96.924\,\kappa_*\,,
  \label{eq:c2Gaussian}
\end{equation}
where we have used the numerical derivative $\widetilde\Delta'(-\pi/2) \approx
3.8565$.

In order to show how the two-body contacts, for the Efimov and the Gaussian
trimers, are related to the shift $\Gamma_0^3$, we use
Eqs.~(\ref{eq:energyfrB}) 
and (\ref{eq:energyfrGB}) to write
\begin{equation}
    \Gamma_0^3= \frac{\text{e}^{-\widetilde\Delta(\xi)/2s_0}}{\cos\xi} 
    - \frac{\text{e}^{-\Delta(\xi)/2s_0}}{\cos\xi}\,.
  \label{eq:gamma}
\end{equation}
We can Taylor expand the right hand side of Eq.~(\ref{eq:gamma}) around
the unitary limit, $\xi=-\pi/2$ obtaining
\begin{equation}
    \Gamma_{0,0}^3=\frac{1}{2s_0}\left(\widetilde\Delta'(-\pi/2)-\Delta'(-\pi/2)\right)\,.
  \label{eq:shiftD}
\end{equation}
Therefore we conclude that the constant term of the shift is proportional to the difference
of the derivatives of the Gaussian and zero-range universal function
at the unitary limit.
The derivatives are shown as the dashed lines in 
Fig.~\ref{fig:uni1}. Combining Eqs.~(\ref{eq:shiftD}), (\ref{eq:c2Universal}),
and (\ref{eq:c2Gaussian}) we obtain
\begin{equation}
  \Gamma_{0,0}^3= \frac{1}{16\pi\kappa_*}(\widetilde C_2 - C_2) \approx
  0.8653\,.
  \label{eq:shiftc2}
\end{equation}

The shift can therefore be related
to a different variation of the three-boson energy with respect to
the scattering length at the unitary limit given by the potential models with
respect to the zero-range case. In the case of real potentials 
moving along the path in which $r_B$ is constant the shift measures the increase
in this variation. Here we have observed almost equal values of the shift using
different potential models with variable strengths to explore the dynamics close to the unitary
limit. However, we point out that in systems governed by a coupled-channel 
dynamics, different values of $\Gamma_{0,0}^3$ are possible, even negative ones.

In the case of the three-body contact, with the definition given in
Eq.(\ref{eq:contact3}) it results $C_3=\kappa_*^2$ in both cases, using the
zero-range theory or the Gaussian model. As before we can related
the latter to the range of the Gaussian as $C_3=0.238/(r^G_0)^2$.

\subsection{Relation with experimental results.}

In atomic trapped systems the measured three-body parameter is  $a_-$ and not
the binding energy at the unitary limit. Therefore the method described before
used to predict the three-body parameter from a known energy
value cannot be applied. However, it is possible to estimate
 the three-body parameter $\kappa_*$ using the value of $a_-$. In fact, 
Eq.~(\ref{eq:energyfrG}) can be solved at $\xi=-\pi$ reducing the strength of
the Gaussian potential up to the point in which the three-body energy results to be zero.
From detailed calculations around
such a point we obtain $\widetilde\Delta(-\pi)\approx -1.83$ and, accordingly,
\begin{equation}
\kappa_*a^-_B=-2.483 \, .
\end{equation}
It should be noticed that the parametrization proposed in Eq.(\ref{eq:expg}) can be used
in connection with Eq.(\ref{eq:gamma}) to approximate $\widetilde\Delta(-\pi)$
using the value of the zero-range universal function,
$\Delta(-\pi)=-0.8262$, derived in
Ref.~\cite{gogolin:2008_phys.rev.lett.} and solving a second order
equation. In such a case we obtain an estimate of $\kappa_* a^-_B\approx -2.44$
in reasonable agreement with the above (more exact) result, showing that the
two-term parametrization of the shift $\Gamma_0^3$ can be extended up to the
three-body threshold into the continuum. Also the constant relation between
the scattering length and the energy scattering length holds at that threshold, 
$a^-_B=a_--r_B$, and it can be used to deduce the three-body parameter in terms of $a^-$.
For systems having a van der Waals tail we propose the following estimate
\begin{equation}
\kappa_*\approx \frac{-2.48}{a_--r_B}\approx\frac{0.22}{\ell}\,,
\label{eq:kappau}
\end{equation}
where we have introduced the van der Waals length $\ell$ and we have approximate
the ratios $a_-/\ell\approx -9.5$ and $r_B/\ell\approx 1.4$. 
The first ratio reflects a class of universality determined by the van der
Waals length (see Ref.~\cite{naidon:2014_physrevA} and references therein) 
whereas the second
ratio can be verified using theoretical models as has been done in the
present work in the case of He or, in the case of Cs, by the model given in
Ref.~\cite{chin:2010_rev.mod.phys.}. 
To be noticed that a very close relation to determine $\kappa_*$ in the case
of two-body Leonnard-Jones potentials has been given in Ref.~\cite{blume2015}
showing that also this kind of potentials are well described using the Gaussian
universal function.

Though approximate, the estimate for the three-body parameter given
in Eq.(\ref{eq:kappau}) agrees well with estimates 
given in the literature. For an helium trimer, using $\ell=5.1\,$a$_0$, 
Eq.(\ref{eq:kappau}) predicts a three-body parameter of $\kappa_*\approx0.043\,$a$_0^{-1}$ 
whereas calculations using the LM2M2 potential give 
$\kappa_*\approx0.044\,$a$_0^{-1}$~\cite{gattobigio:2012_phys.rev.a,hiyama:2014_phys.rev.a}. 
It is interesting to noticed that the above relation is also in agreement
with the estimates of Ref.~\cite{huang:2014_phys.rev.a}
for the three-body parameter of $^6$Li ($\kappa_*=0.00678\,$a$_0^{-1}$)
and $^{133}$Cs ($\kappa_*=0.0017\,$a$_0^{-1}$). In the first case, using
$\ell=31.26\,$a$_0$ Eq.(\ref{eq:kappau}) predicts $\kappa_*\approx
0.007$a$_0^{-1}$, in good agreement with
the estimate. In the second case, using $\ell=101\,$a$_0$ a value of 
$\kappa_*\approx 0.002$a$_0^{-1}$ is obtained, in reasonable agreement with
the quoted value in Ref.~\cite{huang:2014_phys.rev.a}.

\bigskip

\section{Conclusions.}\label{sec:conclusions}

We have analyzed the behavior of a three-boson system close
to the unitary limit using potential models differing in form and scale.
Specifically we have used the LM2M2 helium-helium interaction, the MTIII
interaction constructed to describe the $s$-wave nucleon-nucleon interaction in
triplet state and two Gaussian representations of these potentials.
The strength of the potentials have been varied in order to move the system towards 
the unitary limit. A first observation was that along this path the length 
$r_B=a-a_B$ results to be almost constant. This quantity characterizes the system, 
it is about $7.2\,$a$_0$ for the helium system and about $1.2\,$fm for the nuclear system. 
This can be understood as a particular path to the unitary limit and 
the Gaussian representation of the potential is constructed to
reproduce the length $r_B$ along that path.

The analysis of the three-boson ground-state binding energy close to the unitary limit
using the Gaussian potentials have shown different scaling properties
allowing to determine the three-body parameter, $\kappa^G_*$, from the range of the
Gaussians. Moreover this family of potentials produces a shift with a constant
term of $\Gamma^3_{0,0}\approx 0.87$, independently of the range, and a small
first order correction proportional to $1/\kappa_*a_B$. This characteristic
allows to construct a Gaussian universal function $\widetilde\Delta(\xi)$.
The shift has been shown to be related to the difference between the
zero-range-trimer
two-body contact and the ground-state-trimer two-body contact at the
unitary limit.
Being the contact an observable, this connection allows for
a measurement of the shift.
Finally, the Gaussian universal function $\widetilde\Delta(\xi)$ has been used
to estimate the three-body parameter $\kappa_*$ knowing the two-
and three-boson binding energies at a particular value of the two-body
scattering length $a$. Since in atomic-trapped systems estimates of the
scattering length at dissociation are given, we make the analysis from
the known value of $a_-$ as well. In this case we use the relations
$a_-/\ell\approx -9.5$ and $r_B/\ell\approx 1.4$, 
verified in several atomic species.

In many cases the zero-range universal function $\Delta(\xi)$ has been used
to analyze experimental results with the conclusion that not always a complete 
agreement was found. In those cases in which the agreement is not satisfactory 
we suggest the use of the Gaussian universal function $\widetilde\Delta(\xi)$ 
as an alternative to make the analysis. To this respect a parametrization of the 
function in terms of the angle $\xi$ could be helpful. Studies along this line are 
at present in progress as
well as the analysis of different classes of paths to reach the unitary limit.


%
\end{document}